\documentclass[prl,twocolumn,amsmath,amssymb,superscriptaddress]{revtex4}
\usepackage{graphicx}
\usepackage{dcolumn}
\usepackage{bm}
\begin{document}

\title{Spin-orbit Interaction induced Singlet-Triplet Resonant
Raman Transitions in Quantum-dot Helium}
\author{Aram Manaselyan} \author{Areg Ghazaryan}
\affiliation{Department of Solid State Physics, Yerevan State
University, Yerevan, Armenia}
\author{Tapash Chakraborty}
\affiliation{Department of Physics and Astronomy,
University of Manitoba, Winnipeg, Canada R3T 2N2}

\begin{abstract}
From our theoretical studies of resonant Raman transitions in two-electron
quantum dots (artificial helium atoms) we show that in this system, the
singlet-triplet Raman transitions are allowed (in polarized configuration)
only in the presence of spin-orbit interactions. With an increase of the 
applied magnetic field this transition dominates over the singlet-singlet 
and triplet-triplet transitions. This intriguing effect can therefore be 
utilized to tune Raman transitions as well as the spin-orbit coupling in 
few-electron quantum dots. 
\end{abstract}

\maketitle


Quantum dots (QD) or the artificial atoms \cite{qdbook} containing
two interacting electrons -- popularly known as artificial helium atoms
\cite{helium} have received considerable attention for over a decade
because of their relative simplicity, but at the same time being
rich in fundamental physics. One of the most interesting features in
this system is the spin singlet-triplet transition in a externally
applied magnetic field \cite{sing_trip,ashoori}. This spin transition
is a consequence of an interplay between the electron-electron interaction 
and the harmonic confinement potential \cite{sing_trip}. It has been
proposed that resonant Raman transitions \cite{Koppen,schuller} are
perhaps a direct route to observe this transition. We have studied
Raman scattering in GaAs quantum dot helium for polarized configuration 
and we observed that in accordance with the experimental observations 
\cite{Koppen,Singha,resonant}, due to the polarization selection rules, only 
singlet-singlet transitions are observed for zero magnetic field, while 
triplet-triplet transitions are possible for higher values of the magnetic 
field. In this Letter, we report that in the presence of the Rashba spin-orbit
(SO) interaction \cite{Rashba}, there are additional singlet-triplet
and triplet-singlet Raman transitions that are forbidden without the
SO interaction. Further, the external magnetic field can be used to
tune the amplitudes of these new transitions.

A very useful mechanism for coherent spin manipulation in quantum
nanostructures is via the Rashba SO interaction \cite{Rashba,book} which
couples the orbital motion of electrons with the spin state. The SO 
interaction can arise in a quantum dot due to the confinement and lack 
of inversion symmetry of the nanostructure which creates a local 
electric field perpendicular to the electron plane. The SO coupling 
strength can be varied by changing the asymmetry of the quantum structure 
with an external electric field. There were a few recent reports on the 
tunability of the SO interaction in few-electron quantum dots \cite{nitta}.
In our recent work on the Rashba effects in quantum dots \cite{Pekka,Aram}
we found multiple level crossings and level repulsions in the energy spectrum
that was a result of the interesting interplay between the Zeeman effect and
the SO interaction. The  influence of Rashba and Dresselhaus SO interaction 
on the energy levels and optical absorption spectrum for two-electron QD was 
investigated earlier \cite{Peeters}. We also found \cite{Aram1} that the 
Rashba SO coupling is responsible for additional Raman transitions, the amplitude 
of which can be controlled externally by changing the SO coupling parameter.


Following the experimental work of Singha et al. \cite{Singha}, we consider a 
GaAs/AlGaAs quantum dot with a diameter of 180 nm. We chose the confinement 
potential of the dot as parabolic with an oscillator energy $\hbar\omega^{}_0$. 
The Hamiltonian of the N-electron system in the dot can be written as
\begin{equation}\label{NElectron}
{\cal H}=\sum_i^N{\cal H}_i^e+\frac12 \sum_{i,j}^N
\frac{e^2}{\varepsilon |{\bf r}^{}_i-{\bf r}^{}_j|},
\end{equation}
where the second term describes the Coulomb interaction between electrons, $e$ 
is the electron charge and ${\cal H}_i^e$ is the single-electron Hamiltonian in 
the presence of an external perpendicular magnetic filed and with the SO 
interaction included
\begin{equation}\label{singleH}
{\cal H}_i^e={\frac1{2m^{}_e}}\Pi_i^2+\tfrac12
{m^{}_e\omega_0^2r_i^2}+\tfrac12 g\mu^{}_B B\sigma^{}_z+H^{}_{\rm SO},
\end{equation}
where $\bf \Pi=\bf p-\frac ec \bf A$ and $\bf A$ is the vector
potential of the magnetic field. The third term on the right hand
side of Eq.~(\ref{singleH}) is the Zeeman splitting. The last term
describes the Rashba SO interaction \cite{Rashba}
\begin{equation}\label{Rashba}
H^{}_{\rm SO}=\frac \alpha\hbar \left[{\boldsymbol
\sigma}\times\left({\bf p}-\frac ec {\bf A}\right)\right]^{}_z,
\end{equation}
with $\alpha$ being the spin-orbit coupling constant, which is
sample dependent and is proportional to the interface electric field
that confines the electrons in the $xy$ plane. In (\ref{singleH})
and (\ref{Rashba}), $\boldsymbol \sigma$ is the electron spin
operator and $\sigma^{}_x,\sigma^{}_y$ and $\sigma^{}_z$ are the
Pauli spin matrices. The eigenfunctions of the single-electron
Hamiltonian (\ref{singleH}) can be presented as a linear expansion
of the Fock-Darwin orbitals \cite{qdbook} $f^{}_{n,l}(r,\theta)$,
where $n,l$ are the radial and angular quantum numbers. The Rashba 
term $H^{}_{\rm SO}$ in turn will couple the single-electron state with 
angular momentum $l$, and spin up to the state with angular momentum 
$l+1$, and spin down \cite{Pekka}. The energy spectrum of the many-electron 
system was obtained by diagonalizing the Hamiltonian matrix (\ref{NElectron}).

In order to evaluate the Raman transition amplitudes, first we have to define 
the initial, final and the intermediate states. Let us consider the resonant 
inelastic light-scattering process in a backscattering configuration with the 
incident photon energy just above the effective band gap of the quantum dot and 
with the wave vector transfer in the lateral dimension $q=2\times10^4$cm$^{-1}$
\cite{Singha}. In that case the initial states of the N-electron system will be 
the ground state, and the final states will be the intraband excitations of the 
N-electron system with the same total momentum projection $J^{}_z$ as for the 
initial state. For the intermediate states we have N+1 electrons in the conduction 
band and one additional hole in the valence band. For simplicity, we consider here 
only the heavy-hole states. Under this approximation the single-hole Hamiltonian 
and the wave functions are similar to those for the electron. We need to change 
only the values of the effective mass and the confinement parameter. It is well 
known that the Rashba effect on heavy hole ground state is very weak \cite{Aram}. 
Hence we have neglected the SO effect on the hole states. The hole states can also be
described with the help of the Fock-Darwin functions, and the basis functions of 
the intermediate states can be constructed as products of the Slater determinants 
of the electrons and the single-hole wave functions.

The Raman scattering transition amplitude from the initial state $|i\rangle$ to 
the final state $|f\rangle$ is obtained from \cite{Raman_few}
\begin{equation}\label{Raman}
A^{}_{fi}\sim\sum_{\rm int} \frac {\langle f|H^{(+)}|{\rm int}\rangle 
\langle {\rm int}|H^{(-)}|i\rangle}{\hbar\omega^{}_i-(E^{}_{\rm int}-
E^{}_i)+i\Gamma^{}_{\rm int}},
\end{equation}
where $\hbar\omega^{}_i$ is the incident photon energy. In equation
(\ref{Raman}) $H^{(-)}$ and $H^{(+)}$ are the single-particle
operators describing the photon absorption (-) and emission (+)
processes respectively \cite{Aram1,Raman_few}.

For Raman scattering, we need to consider two cases: (i) the polarized geometry,
i.e., when the polarization vectors of incident and scattered photons are in the 
same direction, and (ii) the depolarized geometry, when the polarization vector 
of the scattered photon is perpendicular to that of the incident one.

The differential cross section of Raman scattering can be calculated using the 
following expression
\begin{equation}
d\sigma\sim\sum_f |A^{}_{fi}|^2\delta(\Delta E-(E^{}_f-E^{}_i)),
\end{equation}
where $\Delta E=\hbar\omega^{}_i-\hbar\omega^{}_s$ is the Raman
energy shift. In our calculations we have used a Lorentzian instead
of the Dirac delta function in order to take into account the level
width of the final states \cite{Aram1}.

\begin{figure}
\includegraphics[width=7cm]{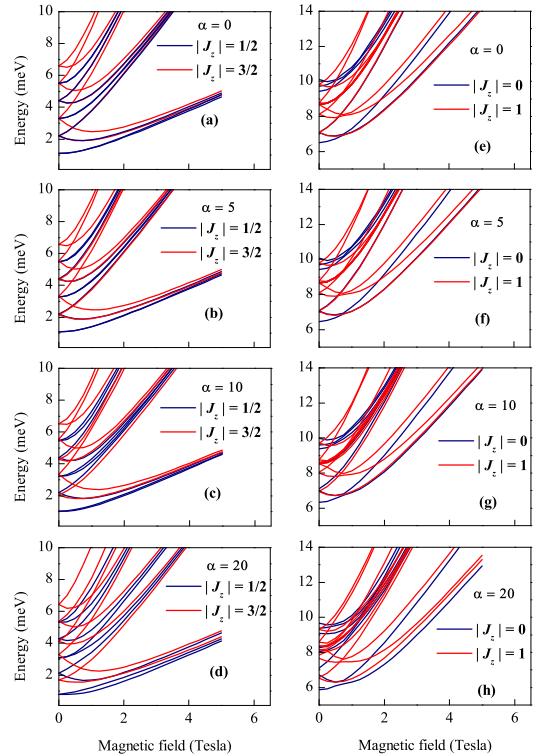}
\caption{\label{fig:EdepB} Magnetic field dependence of the low-lying energy levels 
in a single-electron (a-d) and two-electron (e-h) quantum dot for various values of 
the SO coupling strength $\alpha$ (in meV nm).}
\end{figure}

{\it Energy levels} -- In our present study, we consider a GaAs QD with
$m^{}_e=0.063m^{}_0$, $m^{}_h=0.33m^{}_0$, $\varepsilon=12.9$ and we
used uniform values for level widths $\Gamma^{}_{\rm int}=\Gamma^{}_f=0.5$ meV. 
In Fig.~\ref{fig:EdepB}, the magnetic field dependence of the low-lying energy 
levels of a QD with one and two electrons is presented for different values of 
the total momentum $J^{}_z$ and for four values of the SO coupling parameter 
$\alpha=0,5,10,20$ meV nm. This wide range of values of $\alpha$ provides a clear
dependence of the energy spectra on this parameter. When compared with the Fock-Darwin 
spectra without the SO coupling (Fig.~\ref{fig:EdepB}(a)), the most outstanding features 
in the energy spectra of the quantum dots with SO coupling are the lifting of degeneracy 
at a vanishing magnetic field, rearrangement of some of the levels at small fields, 
and the level repulsion at higher magnetic fields (Fig.~\ref{fig:EdepB}(b) - (d)). 
For zero magnetic field and without the Rashba SO interaction, the ground state of 
a single-electron QD is characterized by $n=0,l=0,\sigma=\pm1/2$ with the
corresponding energy $\hbar\omega^{}_0$. The next two excited states are $n=0,l
=\pm1,\sigma=\pm1/2$ and $n=1,l=0,\sigma=\pm1/2$ with energies $2\hbar\omega^{}_0$ 
and $3\hbar\omega^{}_0$ respectively. The SO coupling in turn mixes the states 
$|l,1/2\rangle$ with $|l+1,-1/2\rangle$ and $|l,-1/2\rangle$ with $|l-1,1/2\rangle$,
which removes the four-fold degeneracy for the first excited state and introduces 
some level repulsions at higher fields. Similar effects are clearly visible also 
for dots with two electrons (Fig.~\ref{fig:EdepB}(e-h)). Therefore, with the SO 
coupling the total angular momentum and the total spin of the electrons are no
longer good quantum numbers and we have to use the total momentum $J^{}_z$ instead, 
to describe the states \cite{Pekka,Aram}.

\begin{figure}
\includegraphics[width=7cm]{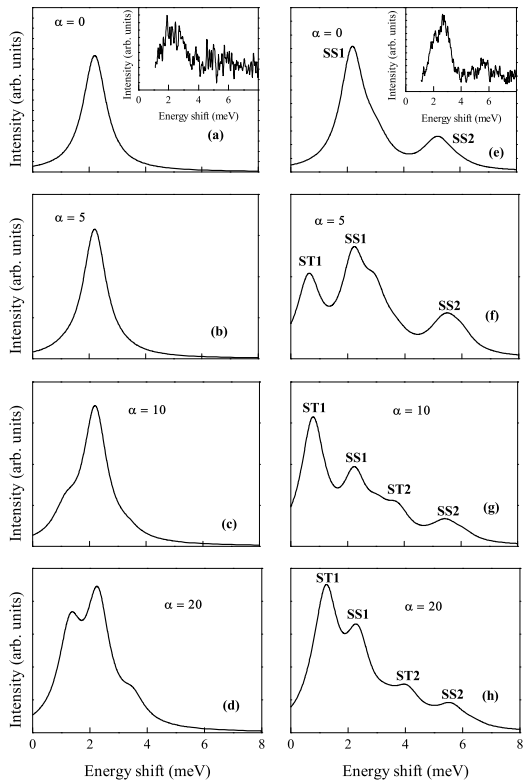}
\caption{\label{fig:RamanRashba0} The Raman scattering amplitudes for the polarized 
geometry and for various values of SO coupling parameter $\alpha$ (in meV nm) for 
single-electron (a-d) and for two-electron (e-h) quantum dot. Insets: Experimental 
data for $\alpha=0$ from Ref.~\cite{Singha}}
\end{figure}

{\it Raman spectra} -- The Rashba SO coupling can play an important role in Raman 
spectroscopy of the QDs. The effect of Rashba SO coupling on the Raman excitations 
for quantum dots containing one electron is shown in Fig.~\ref{fig:RamanRashba0}
(a-d) and for QDs with two electrons in Fig.~\ref{fig:RamanRashba0}(e-h). For a 
single-electron QD we considered the polarized Raman excitations between the states 
with total momentum $J^{}_z=1/2$ (blue lines in Fig.~\ref{fig:EdepB} (a-d)). For a 
two-electron QD we considered similar excitations between the states with total 
momentum $J^{}_z=0$ (blue lines in Fig.~\ref{fig:EdepB} (e-h)).

We begin with the Raman scattering for one and two electron quantum dots without 
the SO coupling (Fig.~\ref{fig:RamanRashba0} (a) and (e)). The experimentally 
measured values of the resonant Raman scattering amplitudes for the polarized 
geometry are presented as insets of Fig.~\ref{fig:RamanRashba0} for a dot with one 
and two electrons. The theoretical results are in good agreement with the
experimental data \cite{Singha}. It is easy to see from the experimental data that 
for the case of the one-electron QD, we have only one peak with the Raman energy 
shift of $\Delta E=2.2$ meV. That peak corresponds to the excitation of the system 
from the ground state $|0,0\rangle$ to the first excited state with the same angular
momentum $|1,0\rangle$ with energy difference $2\hbar\omega^{}_0$. Therefore, we can 
use the value $\hbar\omega^{}_0=1.1$ meV in our calculations for the single-electron 
QD, and we use $\hbar\omega_h=1$ meV for the hole. In the case of a quantum dot 
with two electrons (Fig.~\ref{fig:RamanRashba0}(e) and corresponding inset),
additional Raman modes appear at higher energies. Theoretical studies for the 
two-electron QD are also remarkably similar to the experimental observations. Here 
we have used the value $\hbar\omega^{}_0=1.6$ meV for the two-electron QD, which is 
larger than that for the one electron case \cite{Singha}.

\begin{figure}
\includegraphics[width=7cm]{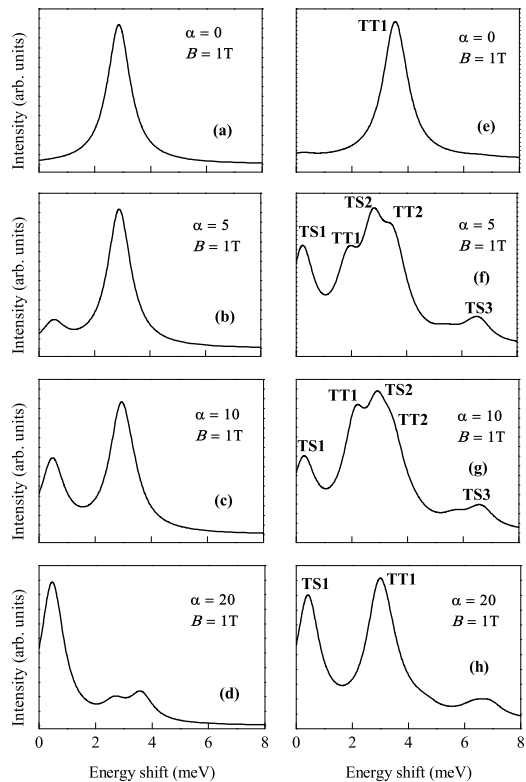}
\caption{\label{fig:RamanRashba1} The same as in Fig.~\ref{fig:RamanRashba0}, but 
for a magnetic field of 1 Tesla.}
\end{figure}

Clearly, the Rashba SO coupling is responsible for additional Raman excitations even 
for the single-electron system. With an increase (or decrease) of the SO parameter 
$\alpha$ it is possible to manipulate the amplitudes of these additional excitations.
To understand this unique effect we consider the first three one-electron states with
total momentum $J^{}_z=1/2$. Raman excitations with higher amplitude are possible 
only between the states with the same angular momentum $l$. Therefore without the SO 
coupling the transition is only from the ground state to the second excited state. 
However, the SO coupling mixes all those states and these can be expressed as linear
combination of states with different angular momenta $l$. Therefore, we now have the 
possibility of Raman transitions from the ground state to both excited states. With 
a further increase of $\alpha$ the weight of $|1,0\rangle$ state in the first excited 
state increases, and so does the transition amplitude. With a decrease of $\alpha$ 
the weight of $|1,0\rangle$ state will vanish and the additional peak will disappear.

\begin{figure}
\includegraphics[width=7cm]{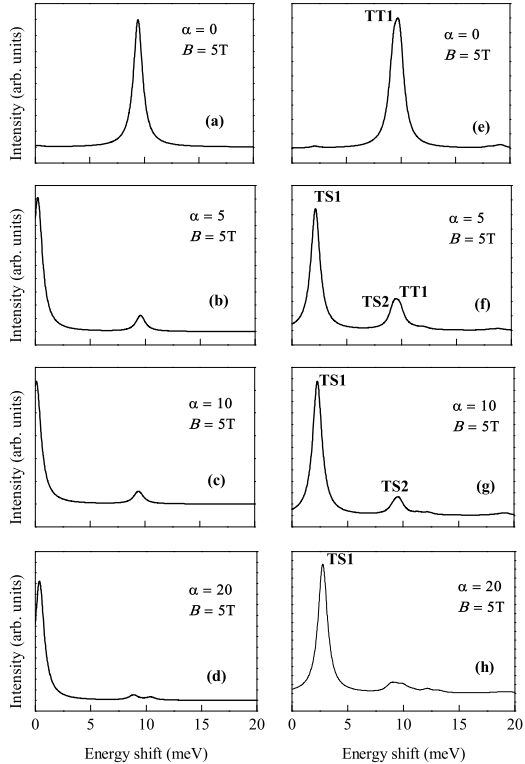}
\caption{\label{fig:RamanRashba5} The same as in
Fig.~\ref{fig:RamanRashba0}, but for a magnetic field of 5 Tesla.}
\end{figure}

A similar situation is also observed for a two-electron QD 
(Fig.~\ref{fig:RamanRashba0}(e)-(h)). Here we see many additional peaks in the 
observable energy range, in comparison to the case of without the SO coupling. In
the latter case and in the absence of the magnetic field the ground two-electron state 
is singlet with $J^{}_z=0$ and the Raman transitions are only to singlet excited states 
with the same total momentum \cite{Koppen}. With the increase of Rashba coupling 
strength $\alpha$ the SO interaction will again mix the states with different total spins 
and therefore we can not characterize the states as fully singlet or triplet. All the 
states must now be presented as superpositions of two-component basis states 
$|n_1,l_1,\sigma_1\rangle|n_2,l_2,\sigma_2\rangle$. The most important component of the 
ground state is the singlet state 
$|0,0,1/2\rangle|0,0,-1/2\rangle$ with weight 70.5\% and we can still characterize 
it as a singlet. Due to the SO mixing, the final states also will have triplet 
components. As an example, the most important component of the final state in the 
first peak in Fig.~\ref{fig:RamanRashba0}(f) is the triplet state $|0,0,1/2\rangle|0,
-1,1/2\rangle$ with weight 86.5\%. Hence we can call that transition as the
singlet-triplet (labelled as ST1) transition. The peaks SS1 and SS2 in 
Fig.~\ref{fig:RamanRashba0}(f) are very similar to the peaks for the case without 
the SO coupling and are essentially singlet-singlet transitions. For these the most 
important components are $|1,0,\pm1/2\rangle|0,0,\mp1/2\rangle$ with weight 86.5\% 
and $|2,0,\pm1/2\rangle|0,0,\mp1/2\rangle$ with weight 69.6\% respectively. In 
Fig.~\ref{fig:RamanRashba0}(f-h) several singlet-triplet (ST) and singlet-singlet 
(SS) transitions are visible. With an increase of the SO coupling parameter $\alpha$ 
the amplitude of the first singlet triplet transition ST1 increases and
it becomes the dominant one.

{\it Magnetic field effect} -- In Figs.~\ref{fig:RamanRashba1}-\ref{fig:RamanRashba5}
we present results as in Fig.~\ref{fig:RamanRashba0} but for various values of the
magnetic field $B$. According to these results, the magnetic field significantly 
changes the Raman spectra. For the single-electron case the peaks have become more 
pronounced and here we again observe the emergence of the transition from the ground 
state to the first excited state by switching on the SO coupling. These figures 
indicate that for $\alpha=20$ the transition amplitude from the ground state to the 
first excited state is several times bigger than the transition to second excited 
state compared to the same figure without the magnetic field.

With an increase of the magnetic field, the two-electron ground state changes from 
singlet to triplet near the field of $B=1$Tesla. Therefore without the SO coupling 
[Fig.~\ref{fig:RamanRashba1}(e)], the most important component of the ground state 
is the triplet state $|0,1,-1/2\rangle|0,0,-1/2\rangle$ with weight 93\% and all
possible transitions are triplet-triplet (TT). With an increase of the SO
coupling $\alpha$, we find additional triplet-singlet (TS) and triplet-triplet 
transitions. For example, the first peak in Fig.~\ref{fig:RamanRashba1}(f) 
corresponds to a transition to the final state with the most important component
$|0,0,1/2\rangle|0,0,-1/2\rangle$ having weight of 70.7\%. Similar to the
single-electron case we can tune the amplitudes of the peaks with the magnetic field,
and again for $B=5$ Tesla (Fig.~\ref{fig:RamanRashba5}) the first additional peak 
created by the SO coupling becomes dominant. It is however important to note that 
for weak magnetic fields that peak can be characterized as a singlet-triplet, and for 
higher values of the magnetic field it becomes triplet-singlet.

To summarize, we have studied the influence of the Rashba spin-orbit coupling on 
the resonant Raman electronic excitations in one- and two-electron GaAs quantum dots 
for polarized configuration. We have shown that the SO coupling brings in additional 
Raman transitions, the amplitudes of which depends on the coupling parameter $\alpha$. 
In the case of a two-electron QD, in addition to the usual singlet-singlet and 
triplet-triplet Raman transitions we also observe the singlet-triplet and 
triplet-singlet Raman transitions. The external magnetic field can be used to tune 
the amplitudes of Raman transitions for both one and two-electron systems.

The work was supported by the Canada Research Chairs Program of the
Government of Canada and Armenian State Committee of Science (Project No.
11B-1c039). The authors are grateful to Vitorio Pellegrini for providing
us with the experimental data of \cite{Singha}.

\end{document}